\documentclass[prl,aps,superscriptaddress,showpacs,showkeys,reprint]{revtex4-1}
\usepackage{amsmath,amsfonts,amssymb}
\usepackage{bm}
\usepackage{natbib}
\usepackage{graphicx}
\usepackage{array}
\usepackage{verbatim}
\usepackage{multirow}
\usepackage{url}
\usepackage{tikz}
\usepackage[nice]{nicefrac}
\usepackage{hyperref}
\usepackage[mathscr]{euscript}

    \let\e=\varepsilon
   \let\k=\kappa
     
\let\s=\sigma   

    \let\L=\Lambda

\def\nn{\nonumber}

\def\be{\begin{equation}}
\def\ee{\end{equation}}
\def\bea{\begin{eqnarray}}
\def\eea{\end{eqnarray}}
\def\ba{\begin{array}}
\def\ea{\end{array}}
\def\L{\mathcal{L}}

\begin{document}

\title{Magnetoresistance of an Anderson insulator of bosons}
\author{Anirban Gangopadhyay}
\affiliation{Center for Nanophysics and Advanced Materials, Department of Physics,University of Maryland, College Park, Maryland 20742-4111, USA}
\author{Victor Galitski}
\affiliation{Center for Nanophysics and Advanced Materials, Department of Physics,University of Maryland, College Park, Maryland 20742-4111, USA}
\affiliation{Joint Quantum Institute, Department of Physics, University of Maryland, College Park, Maryland 20742-4111, USA}

\author{Markus M\"uller}

\affiliation{The Abdus Salam International Centre for Theoretical Physics, P. O. Box 586, 34151 Trieste, Italy}
\affiliation{Kavli Institute for Theoretical Physics, University of California Santa Barbara, CA 93106-4030}

\pacs{73.50.Jt, 74.81.Bd, 05.30.Jp, 72.20.Ee, 71.55.Jv}

\begin{abstract}
We study the magnetoresistance of two-dimensional bosonic Anderson insulators. We describe the change in spatial decay of localized excitations in response to a magnetic field, which is given by an interference sum over alternative tunnelling trajectories. The excitations become more localized with increasing field (in sharp contrast to generic fermionic excitations which get weakly delocalized): the localization length $\xi(B)$ is found to change as $\xi^{-1}(B)-\xi^{-1}(0)\sim B^{4/5}$. The quantum interference problem maps onto the classical statistical mechanics of directed polymers in random media (DPRM). We explain the observed scaling using a simplified droplet model which incorporates the non-trivial DPRM exponents. Our results have implications for a variety of experiments on magnetic-field-tuned superconductor-to-insulator transitions observed in disordered films, granular superconductors, and Josephson junction arrays, as well as for cold atoms in artificial gauge fields.
 \end{abstract}

\maketitle

 Transport in Anderson insulators \citep{Anderson1958, ReviewBook50years} is crucially determined by the properties of localized wavefunctions.  Their structure is very complex, both deep in the insulator, as well as upon approaching the delocalization transition, where  they develop a multifractal structure~\cite{MirlinRMP}. A particularly important tool in probing  the nontrivial structure of localized states in Anderson insulators  is magnetoresistance. This is because a magnetic field sensitively affects the quantum interference which in turn influences quantum localization. This effect of the magnetic field has been studied extensively in the past concentrating mostly on non-interacting fermions~\cite{Nguen1985,*Shklovskii1991,KardarBook}. 

Recent experiments on disordered superconducting films provide evidence for {\em bosonic} insulators with localized electron pairs as carriers~\cite{Sacepe2011,Gantmakher2011}. These and other similar systems feature a giant peak in magnetoresistance (MR)~\cite{Sambandhamurthy2004,Paalanen1992,Baturina2008, Goldman2011, Kapitulnik2005}.  This is often interpreted as  a crossover from bosonic to fermionic transport~\cite{Mitchell2011,Skinner2012}, even though the details remain controversial.  Bosonic localization problems arise also in disordered granular superconductors in the insulating regime, in cold bosonic atoms in speckle potentials (where artificial gauge fields can mimic a magnetic field) as well as in disordered quantum magnets. 

The predominant mode of transport in disordered insulators is variable-range hopping of carriers between localized excited states~\cite{ES1984}. The spatial decay of wave-functions describing these localized excitations determines the inelastic hopping rate and thus the resistance. At low temperature, the (phonon-assisted) hops become significantly longer than the average distance between impurity sites hosting the excitations. 
In this situation, one needs to know the wave-function amplitudes at distances greater than the Bohr radius of an impurity state. At these distances, the amplitude is reinforced by multiple scatterings from  intermediate impurities whereby many alternative paths interfere with each other~\cite{Nguen1985, KardarBook}. 

A perpendicular magnetic field affects the interference of the scattering paths on all length scales and modifies the localization properties.  
Interestingly, bosons and fermions behave very differently in this respect: while in the absence of a field fermion paths typically come with amplitudes  of arbitrary signs, low energy bosonic amplitudes are positive and thus interfere in a maximally constructive way. The magnetic field suppresses this interference, yielding a strong positive magnetoresistance. It exceeds by far a largely opposite effect seen in fermions, which arises from a subtle suppression of negative interferences~\citep{Muller2011}. 

Despite numerous studies of fermionic MR~\cite{Medina1992,Nguen1985,Zhao1991,Imry1988}, a full understanding of the effect of magnetic field on the large-scale structure of localized wave-functions has not been obtained. In this Letter we study the bosonic cousin of this  problem and show that it is amenable to a complete solution. The simplifying circumstance is the absence of  additional sign-factors in the latter quantum interference problem, which allows a mapping to classical statistical mechanics of directed polymers in random media (DPRM). More generally, our analysis of MR is also valid for fermionic problems, provided the interfering paths have essentially only positive amplitudes. This arises, e.g., in the tunneling through the bottom of the conduction band in a solid semiconductor solution~\cite{Shklovskii1982}, or in fermionic impurity bands with Fermi level very close to the band bottom~\footnote{In the impurity band model considered below, the distance of the Fermi level from the bottom of the band should be $\lesssim 10\%$ of the bandwidth for positive MR to occur in some range of finite $B$. At smallest $B$, MR of fermions is almost always negative, however~\cite{Nguen1985,*Shklovskii1991}.}.  

{\em The model --} Here we study a model of hard-core bosons on a square lattice,
\begin{equation}
\label{Hamiltonian}
 H = \sum_i (\e_i-\mu) c_i^\dagger c_i - t \sum_{\left\langle ij \right\rangle} \exp\left[ i \int_{{\bf r}_i}^{{\bf r}_j} d\mathbf{r}\cdot\mathbf{A}\right] c_i^\dagger c_j+{\rm h.c.},
\end{equation}
with uniformly distributed on-site disorder
in the range $\e_i \in [-W,W]$. We take $W=1$ as the energy unit and consider weak nearest-neighbor tunneling, $t\ll W$. We fix the chemical potential to $\mu=0$ to study a half-filled impurity band. A perpendicular magnetic field is introduced via the vector potential $\mathbf{A}=Bx\, \mathbf{e_y}$, with $B$ being the flux per plaquette in units of the flux quantum. 
\begin{figure}
 \centering
 
 \includegraphics[angle = 0,width = 0.47 \textwidth]{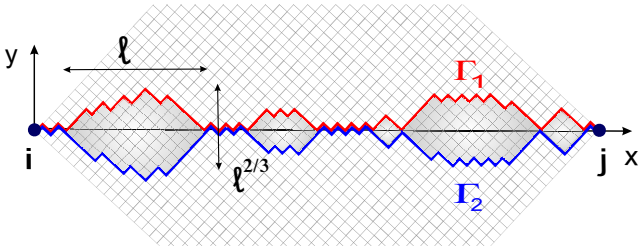}
 
  \caption{ 
The approximation of directed propagation~\cite{Nguen1985,*Shklovskii1991} maps the wavefunction 
to a directed polymer. The droplet picture suggests that traces of localized wavefunctions, or low energy polymer configurations, form a string of  loops of competing/interfering paths.  Relevant loops of size $\ell$ have transverse roughness $\sim \ell^{\zeta=2/3}$). They are rare, being separated by a typical distance $\ell^{1+\theta}=\ell^{2\zeta}\gg \ell$. Two competing paths $\Gamma_{1,2}$ are shown, and the loops/droplets they form.}
 \vspace*{-0.1 in}
 \label{InterferenceSum}
\end{figure}

We now focus on the spatial structure of an excitation localized around site $i$. It is characterized by the residue of the pole at $\omega\!\approx\! \epsilon_i$ of the retarded Green's function $G^R_{j,i}(\omega) = -i\int_0^\infty dt e^{i\omega t} \langle [c_j(t),c_i^\dagger(0)] \rangle$ \footnote{This follows immediately from the Lehmann representation of the Green's function}. Its decay away from the site $i$ defines a localization length. Deep in the insulating regime, $G^R_{j,i}$ can be evaluated using a locator expansion \cite{Muller2011}. To leading order in small hopping one obtains a sum over all paths $\Gamma$ of shortest length~\cite{Nguen1985, *Shklovskii1991}, ${\rm dist}(ij)$ (cf. Fig. \ref{InterferenceSum}: only  right-going steps are allowed)
\bea
\label{result}
S_{ji}(B)\equiv \left. \frac{1}{t^{{\rm dist}(ij)}}\frac{G^R_{j,i}(\omega)}{G^R_{i,i}(\omega)} \right|_{\omega \to \e_i}\!\!\!\!\!\! = \sum_\Gamma e^{i\Phi_\Gamma(B)} J_{\Gamma}(\omega=\e_i),\quad\,
\eea
which is closely analogous to the sum over paths for fermionic Anderson insulators~\cite{Anderson1958}.
In Eq.~(\ref{result}) each path $\Gamma$ contributes with an amplitude 
\bea
\label{amplitude}
J_{\Gamma}(\omega) = \prod_{k\in \Gamma\setminus  \{i\}} \frac{\mbox{sgn}(\e_k)}{ \e_k-\omega}.
 \eea 
and an accumulated phase $\Phi_\Gamma(B) = \int_\Gamma d\mathbf{r}\cdot\mathbf{A}$. 
On average, the larger the excitation energy $\e_i$, the faster the spatial decay of $\left| S_{ji}\right|$ ~\citep{Muller2011}. Henceforth, we focus on low-frequency excitations (relevant for transport at low $T$) and hence set $\omega=\e_i=0$. 

\begin{figure}
 \centering
 \includegraphics[angle = 270,width = 0.48 \textwidth]{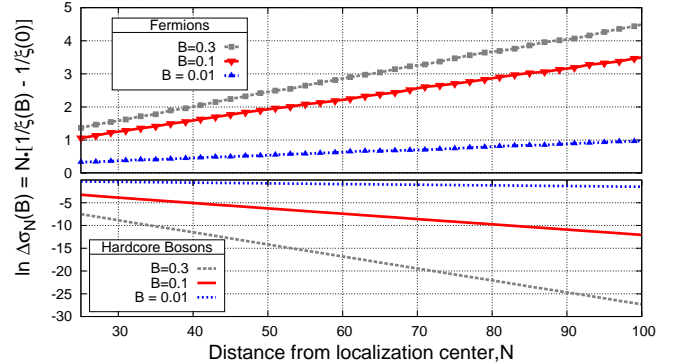}
 \vspace*{-0.2 in}
 \caption{Magnetoconductance of fermions and bosons as a function of distance $N$ in a half filled impurity band ($\mu=0$). The linear dependence implies that the magnetic flux $B$ changes the localization length $\xi$. While it increases slightly for fermions, it shrinks much more substantially in bosons. 
 }
 \label{ConstructiveVsDestructiveInterference}
 \end{figure}

Within this ``forward-scattering approximation"~\cite{Nguen1985, *Shklovskii1991}, justified for $t\ll W$, bosons and fermions differ only by the presence and absence (respectively) of the factor $\mbox{sgn}(\e_k)$ in the amplitudes (\ref{amplitude}). For bosons, the amplitudes are all positive for $\e_i\!=\!0$. A magnetic field destroys this complete constructive interference, and thus localizes the wavefunction more~\cite{Zhao1991, Muller2011, Syzranov2012}.
In contrast, typical fermionic problems~\citep{Nguen1985, *Shklovskii1991} feature amplitudes which vary in sign, depending on the number of sites on the path with $\e_i<\mu$ which are occupied in the ground state. In this case the dominant effect of a magnetic field lies in destroying negative interferences of competing paths, which tends to delocalize the wave function slightly.
Both cases are readily amenable to efficient numerical studies via transfer matrices~\cite{Nguen1985, *Shklovskii1991,Medina1992}, which we use below.
The results shown in Fig.~\ref{ConstructiveVsDestructiveInterference} illustrate the opposite trends.


The relevant quantity for transport is the {\em typical} spatial decay  of localized excitations. Therefore one focuses on the (typical) magnetoconductance, defined as~\cite{Nguen1985, *Shklovskii1991}
\bea
\label{MCdef}
\Delta\sigma_N(B) = \exp\left(\overline{\ln[\left | \nicefrac{ S{ji}(B)}{ S_{ji}(0) } \right|]}\right), \, N\equiv{\rm dist}(ij),\,\,\,\,\,\,
\eea
where the overbar denotes the disorder average. We take $(i,j)$ on opposite corners of a square~\footnote{This comes closest to the situation of more realistic disordered lattices where the disorder average is isotropic.~\cite{Nguen1985,*Shklovskii1991} Note that the Hamming distance $N$ corresponds to the Euclidean distance measured in units of one half of the plaquette diagonal.} (cf.  Fig.~\ref{InterferenceSum}).  
The linear variation with distance in Fig.~\ref{ConstructiveVsDestructiveInterference} implies that at large scales $B$ changes the typical decay rate, i.e., the inverse localization length $1/\xi$, of the excitations. 

\begin{figure}
 \centering
 \includegraphics[angle = 0,width = 0.48 \textwidth]{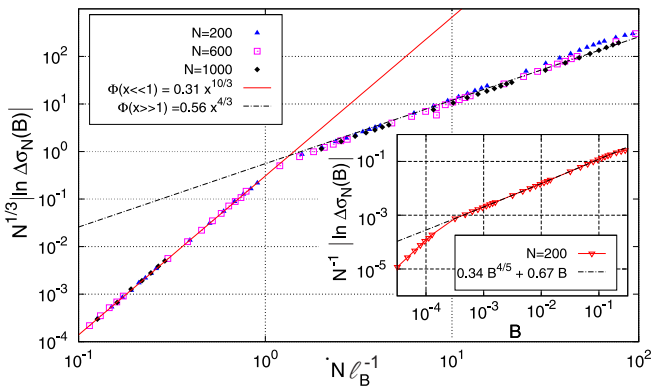}
 \vspace*{- 0.3 in.}
 \caption{Scaling of the magnetoconductance, $\Delta\sigma$, with distance $N$ and flux per plaquette, $B$. The crossover from the perturbative regime $\left|\ln\Delta\sigma_N(B) \right|\sim B^2N^3$ to the non-perturbative regime $\left| \ln\Delta\s_N(B) \right| \sim NB^{4/5}$ occurs at $N\sim\ell_B$, where many successive interfering loops start contributing. Inset: change of inverse localization length for $N=200$, and best fit to the leading two terms in Eq.~(\ref{VirialExpansionMain}), $\xi^{-1}(B)-\xi^{-1}(0)=c_1 B^{4/5} + c_2 B$. 
 }
 \label{BosonsLocLenVsB}
\end{figure}

\emph{Numerical evaluation - }
One numerically evaluates $ S_{ji}(B) \equiv S_{x_j,y_j}(B)$ (with $i$ as origin) by recursion 
\bea
\label{recursion}
\!\!\!\!\!S_{x+1,y}(B) = V_{x+1,y} \left[ e^{i\phi_{-}} S_{x,y-1}(B) + e^{i\phi_{+}} S_{x,y+1}(B)\right]
\eea
 with $\phi_{\pm} = \int_{\Gamma_{\pm}} \mathbf{A} \cdot d\mathbf{r}$, where $\Gamma_{\pm}:(x,y \pm 1) \rightarrow (x+1,y)$ are straight paths along the lattice links and $V_{x,y} = \nicefrac{1}{\left| \e_{x,y}\right|}$. $\Delta \sigma_N(B)$ evaluated from this varies as $B^2 N^3$ for small $(B,N) $ and shows a sharp crossover to $NB^{4/5}$ at larger fields/distances (cf.~Fig.~\ref{BosonsLocLenVsB}). The data for different $N$ is found to collapse onto a scaling function
\bea
\label{crossover}
\left| \ln \Delta\sigma_N(B) \right| &=& N^{-\nicefrac{1}{3}} \Phi\left(NB^{\nicefrac{3}{5}}\right),\\
\Phi(x\ll 1) = b_1 x^{\nicefrac{10}{3}}\,&;&\, \Phi(x\gg 1) = b_2 x^{\nicefrac{4}{3}}, \nn
\eea
with $b_1\approx 0.31, b_2\approx 0.56$. This scaling is expected theoretically from the physics of directed polymers (DPRM), as we explain below.

\emph{Mapping to directed polymers -}
By virtue of the positive path amplitudes $S_{ji}(B=0)$  can be interpreted as the partition sum of a DPRM in 1+1 dimensions~\cite{Huse1985,Halpin-Healy1995}  with random onsite energies $\ln|\e_i|$ (at temperature $T=1$) and ends fixed at sites $i$ and $j$. Each polymer configuration corresponds to a directed path $\Gamma$ of the expansion (\ref{result}).

In low dimensions, DPRM exhibit a pinned phase at large scales, as the random potential is relevant under renormalization~\citep{Larkin1979, Fisher1991}. Beyond a characteristic pinning scale $L_c$ (of the order of the lattice scale here), the random potential competes strongly with the polymer's entropic elasticity and induces roughness exceeding that of random walks: On longitudinal scales $\ell$, typical transverse excursions of configurations grow as $ \ell^\zeta$ with $\zeta > 1/2$.
A low energy excitation that differs from dominant configurations on scale $\ell$, has typical excitation energy $E(\ell) \sim \ell^\theta$, with energy exponent $\theta = 2 \zeta - 1$~\cite{Hwa1994}. In 1+1 dimensions (MR in 2d), the value $\zeta = 2/3$ is known exactly~\cite{Huse1985a}, while $\zeta_{3d}\approx 0.62$ is known numerically~\cite{Tang1992}. 

When $B\neq 0$, the polymer configurations acquire complex weights. 
Studies of $\zeta$ and $\theta$ exponents of complex DPRM~\cite{Medina1989} suggest that the scalings of the pinned phase do not change with complex weights. In fact, for fermions at $B=0$, where negative weights are abundant, there is numerical evidence that the wavefunctions are still governed by DPRM exponents
~\cite{Prior2005,Somoza2009,Monthus2012a}. We thus assume that the DPRM exponents hold for finite fields as well.

It is interesting to note that for weak fields, Eq.~(\ref{recursion}) admits
a continuum limit, where $S$ obeys the equation
\bea
D_xS = D_y^2 S + V(x,y)S,
\label{KPZ}
\eea
with a $\delta$-correlated random potential term $V(x,y)$ and $D_{\alpha=(x,y)} \equiv \partial_\alpha - iA_\alpha(x,y)$ being the gauge-covariant derivative (in Landau gauge $A_y=0$). 
 This generalizes the Kardar-Parisi-Zhang (KPZ) equation~\cite{Kardar1986} 
 to the presence of  complex potentials $V\to V+iA_x$, and may render bosonic MR amenable to a field theoretic analysis similar to Refs.~\cite{FosterNelson,Frey1994}. However, a rigorous study of this modified KPZ equation is not attempted here.

In DPRM language, the magnetoconductance can be cast as a thermodynamic average of the phase factors $e^{i \Phi_\Gamma(B)}$ over polymer configurations,
and the ratio of amplitudes $S_{ji}$ takes the manifestly gauge-invariant form: 
 \bea
 \label{TunnelingProbability}
\left|\frac{S_{ji}(B)}{S_{ji}(0)} \right|^2
= \left[ \frac{\sum\limits_{\Gamma,\Gamma'}e^{-E_\Gamma - E_{\Gamma'}} \cos \left( BA_{\Gamma\Gamma'}\right)}{\sum\limits_{\Gamma,\Gamma'}e^{-E_\Gamma - E_{\Gamma'}}} \right].
 \eea
Here $ E_\Gamma = \sum_{k \in \Gamma \setminus i} \ln \left| \e_k \right|$ is the energy of configuration $\Gamma$, and $A_{\Gamma\Gamma'}$ is the oriented area enclosed by $\Gamma$ and $\Gamma'$.
    
  
  
\emph{MR in weak fields -}  
For weak fields or short distances one can evaluate $\Delta\sigma_N(B)$ perturbatively in $B$. Typical loops of linear extent $\ell$ enclose a flux $\sim B \ell^{1+\zeta}$. Of the $N/\ell$ possible independent loops only a fraction $\sim \ell^{-\theta}$ interfere significantly, cf. Fig \ref{InterferenceSum}, and are thus sensibly affected by $B$. As long as $N\ll \ell_B\equiv B^{-\frac{1}{\zeta + 1}}$ the dominant contribution to Eqn. \ref{TunnelingProbability} comes from the largest loops of length $\ell\sim \!\!N$, which nevertheless enclose only  a fraction of a flux quantum. This results in the magnetoconductance (\ref{MCdef}) $\Delta \sigma_N\propto - N^{- \theta} (B N^{1+\zeta} )^2=-B^2 N^3$. Note that the roughness exponent drops out of this perturbative result. We therefore recover the same scaling as previous authors predicted for interfering paths with positive weights~\cite{Nguen1985, *Shklovskii1991}, even though they assumed random walk scaling, $\zeta=1/2$. However, this coincidence hides the fact that typical wavefunctions are less strongly affected by $B$ than might be suggested by $\Delta \sigma_N$, since the disorder average is dominated by rare events.

\emph{MR in strong fields -}
For $ N\!>\!\ell_B$, DPRM scalings show more clearly in the magnetoresponse. The dominant contribution to $\Delta \sigma_N$ comes from reduced interference in loops of length $\ell_B$, each of which decreases $\Delta \sigma_N$ by $O(1)$.
Larger loops contribute similarly, but 
their probability to interfere significantly decreases as $\ell^{-\theta}$. On the other hand, smaller loops, albeit more abundant and likely to interfere, enclose a small fraction of a flux quantum, and thus have a negligible effect.
The contribution from loops of size $\ell_B$ gives rise to an extensive $\ln(\Delta\sigma_N)$ proportional to the density of significantly interfering loops, 
\bea
\label{nonpertresult}
\frac{\ln \Delta \sigma_N}{N} \equiv -\Delta\left(\frac{1}{\xi}\right) \sim -\ell_B^{-1}\ell_B^{-\theta}= - B^{\frac{1+\theta}{1+\zeta}}= - B^{\frac{2\zeta}{1+\zeta}}.\quad
\eea
This is equivalent to a reduction of the inverse localization length by $B^{4/5}$ in 2d. In $3d$ the same arguments apply, with an exponent ${2\zeta}/(1+\zeta)\approx 0.765$.  Both exceed the value $2/3$ obtained upon neglecting pinning and assuming random walk scaling with $\zeta=1/2$ \cite{Nguen1985, *Shklovskii1991,Shklovskii1983,Zhao1991}.



\begin{figure}
 \centering
 \includegraphics[width=0.45\textwidth]{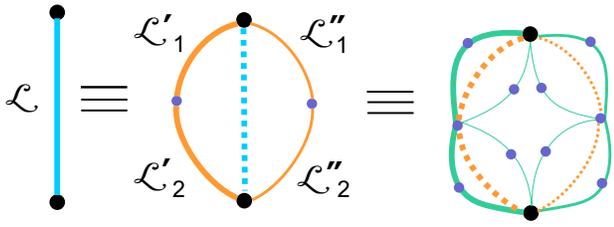}
 \caption{Hierarchical droplet model: At each level of the hierarchy, a parent loop $\L$ (composed of a dominant and subdominant branch) is split into four subloops, two forming the dominant branch ($\L'_{1,2}$, thicker line), and two forming the subdominant branch ($\L''_{1,2}$, thinner lines), cf. Eq.~(\ref{MKRecursion}).  The parent levels are indicated by dashed lines. The dots indicate the splitting into two successive loops at the next level. }
 \label{MKConstruction}
\end{figure}

So far we have discussed the leading scaling with magnetic field. However, the numerical data show small subleading corrections (cf. inset of Fig.~\ref{BosonsLocLenVsB}). Those are indeed to be expected from spatially overlapping loops. To understand their effect, we introduce a hierarchical model which incorporates the essential ideas of droplet theory for directed polymers~\cite{Fisher1991, Hwa1994}. At a given length scale $L$, the polymer  has typically a preferred set of configurations, which compete with alternative, subdominant sets of paths. The leading subdominant family of paths has a higher free energy by $O(L^\theta)$ and wanders off the dominant configuration by $L^\zeta$, enclosing a typical loop area $O(L^{1+\zeta})$. This pattern repeats at all length scales. We simplify this phenomenology by considering a model where loops and alternative paths are restricted to lengths $L_k= N 2^{-k}$ where $N\gg 1$ is the fixed distance between endpoints. Each parent loop of size $L_k$ is composed of a dominant and a subdominant set of paths, each being made up of two successive loops of size $L_{k+1}$,  cf.~Fig. \ref{MKConstruction}. We define the propagation amplitude over the distance $N$ recursively. For a parent loop $\L$ at level $k$ we encapsulate DP scaling by defining the amplitude     
 \bea
\label{MKRecursion}
 S^k_{\L} = S^{k+1}_{\L^{'}_1} S^{k+1}_{\L^{'}_2}+ e^{-f_\L L_k^\theta} e^{ia_\L B L_k^{1+\zeta}} S^{k+1}_{\L''_1} S^{k+1}_{\L''_2},
\eea
where $\L^{'}_{1,2} $ and $\L''_{1,2}$ are the child loops along the dominant and the subdominant path, resp. $f_\L>0$ and $a_\L$ are random variables of order $O(1)$, with a probability density $\rho(f_\L,a_\L)$, assumed to be i.i.d. for all loops $\L$. The recursion is closed by setting all $S^k_{\L} =1$ for $k$ with $L_k\lesssim \ell_B$~\footnote{See the Supplement, Sec. IA, for a discussion of the short scale cut-off, and an alternative definition of the hierarchical model with no restriction on the loop lengths.}. 
The magnetoresistance is defined as $\Delta\sigma_N=\overline{\ln(|S_{0N}(B)/S_{0N}(0)|)}$.

This model has elements in common with the hierarchical lattices analyzed in Ref.~\onlinecite{Derrida1989}. However, here we  {\em explicitly} include the known scaling of excitation energies and areas of loops. The latter is necessary to discuss physically meaningful magnetoresponse. Note that significant interference between the paths $\L'$ and $\L''$ (as given in Eqn. \ref{MKRecursion}) occurs only for rare `active loops' $\L$ for which $f_{\L} L^\theta \lesssim 1$.  
 
The perturbative scaling $\Delta\sigma_N\sim B^2N^3$ is easy to obtain in this model~\footnote{See Sec. II, Supplementary material for details.}. In the non-perturbative regime ($N\gg \ell_B\gg 1$), using that active loops are sparse, one can expand $\ln \Delta\sigma_N(B)$ in powers of the density of active loops of linear size $\ell_B$,~\footnote{See Sec.~IB, Supplementary Material, for details.}
 

\begin{equation}
 \label{VirialExpansionMain}
 \frac{  \ln \Delta\sigma_N(B)}{N} = -B^{\frac{2\zeta}{1+\zeta}}[c_1  + c_2 B^{\frac{\theta}{1+\zeta}} + c_3 B^{\frac{2\theta}{1+\zeta}}  + \cdots ],
 \end{equation}
where the constants $c_i$ depend only on the distribution $\rho(f,a)$. Similar to the Mayer cluster expansion for a system of interacting particles, one obtains a term of $O \left( B^{\nicefrac{(1+n\theta)}{(1+\zeta)}}\right)$ by collecting contributions with exactly $n$ active loops. The leading coefficient $c_1$ is positive definite, and we found $c_2>0$, independently of our choice of the distribution $\rho(f,a)$. Subleading terms due to interfering loops thus enhance the negative MR of bosons. This may explain a similar effect seen in the numerical data on the original lattice (inset of Fig.~\ref{BosonsLocLenVsB}), where a fit yields $c_1\approx 0.34, c_2\approx 0.67$. Hence, $\ln\Delta\sigma$ appears to follow a power law with slightly larger exponent than $4/5$.

\emph{Experimental consequences - }
 In transport through variable-range hopping, at fixed $T$, the resistance depends on the localization length as $R(\xi)=\rho \exp \left(A/\xi^\alpha\right)$,
 with $\alpha=\nicefrac{1}{2}$ (with Coulomb gap) and $\nicefrac{2}{3}$ without (Mott's law in $d=2$)~\cite{ES1984}. According to (\ref{nonpertresult}) a perpendicular magnetic field reduces the bosons' localization length $\xi$ as
 \bea
 1/\xi(B) \approx 1/\xi + \Delta[1/\xi](B),\quad B_{\rm min} \lesssim  B.
 \eea
where $B_{\rm min}= [\xi \ln(R(0)/\rho)]^{-5/3}$ is needed for $\ell_B$ to be shorter than the typical hopping distance. 
To lowest order this effect increases  the resistance by the factor
 \bea
R(B)/R(0)=\left[R(0)/\rho\right]^{ \alpha\, \xi \Delta[1/\xi]}, \quad B_{\rm min} \lesssim  B.\quad
 \label{ResistanceChange}
 \eea
For $B\ll 1$, the exponent is $\alpha c_1 \xi B^{4/5}$. For $B\to 1$, it receives subleading enhancements, reaching values as big as $0.3 \alpha \xi$ \cite{footnote}, cf. Fig.~\ref{BosonsLocLenVsB}. As resistances up to $R(0)/\rho\sim 10^{6}$ are  measurable, and localization lengths $\xi\lesssim 2$ are expected to be within the regime of applicability of forward scattering (loops being sufficiently suppressed) our theory predicts {\em strongly positive} MR of bosons, with enhancement factors of up to two orders of magnitude, within the theoretical and experimental limits. These effects are even stronger when resonances are suppressed~\cite{footnote}. A further enhancement of MR is expected in the critical regime where loops must be included. In contrast, the analogous fermionic problem exhibits {\em negative} MR, which moreover reaches a much smaller maximal amplitude, cf. Fig.~\ref{ConstructiveVsDestructiveInterference}. The importance of the bosonic MR makes it likely to be a key ingredient in the MR peak observed in superconducting films with preformed pairs~\cite{Sambandhamurthy2004}. Finally, it would be interesting to probe for the predicted magnteoresponse and its sensitivity on quantum statistics using cold atoms subjected to artifical gauge fields.


 We would like to thank A. Dobrinevski, P. Le Doussal and B.I. Shklovskii for very useful discussions. This research was supported by NSF  DMR-0847224 (A.G.) and DOE-BES DESC0001911 (V. G.) and  NSF-KITP-12-183 (M.M.).

  \bibliography{LocLenVsB}

\begin{thebibliography}{50}
\bibitem [{foo()}]{footnote}%
  \BibitemOpen
  \href@noop {} {\bibinfo  {journal} {Upon excluding resonances by constraining
  $\epsilon_i \in [-1,-1/2] \cup [1/2,1]$, positive MR is enhanced, and the
  exponent in Eq.~(\ref{ResistanceChange} can reach values up to $\sim 0.6
  \alpha \xi$}\ }\BibitemShut {NoStop}%
\end{thebibliography}
  \begin{widetext}
  
  \end{widetext}
  
\newpage

  \section*{Supplementary material for Bosonic Anderson insulators in a magnetic field}

\section*{I. HIERARCHICAL LOOP MODEL}

Here we analyze in more detail the hierarchical loop model defined in the main text. This model implements the ideas of the droplet picture in an analytically tractable and mathematically precise way. Our focus will be on the analytical calculation of magnetoconductance in the perturbative and the non-perturbative regimes. However, we have also studied the crossover between the two regimes numerically. We found the crossover, the asymptotic power laws and subleading corrections to be very similar to those observed in the full lattice model of forward-directed paths. This suggests that the hierarchical droplet model captures indeed most of the relevant physical ingredients of magnetoconductance.

\subsection{A. Models}

Imposing the scaling of individual droplet degrees of freedom  actually does not fully specify a hierarchical droplet description, but leaves some freedom in the definition of the model. The resulting models differ in the way they treat correlations between energies of spatially overlapping droplets. As we will see this translates primarily into differences in the numerical coefficients of subleading terms.

\subsubsection{1. Normalized recursion}
We first discuss a different version of the hierarchical construction from the one in the main text.
We define it by iterating the following recursive construction from the largest scale $N$ down to the lattice scale,
the loops or branch segments having lengths $L_k = 2^{-k}N$ for $0 \leq k\leq  K\equiv \lfloor \log_2(N) \rfloor$,
\bea
S_{\L}&=&1, \quad {\rm if}\quad L_\L= 2^{-K}N,\nn\\
S_{\L}&=& \frac{S_{\L'_1}S_{\L'_2}+W_\L(B)  S_{\L''_1}S_{\L''_2}}{1+W_\L(0)},\label{construction}\\
W_\L(B) &=& \exp\left[-f L_\L^\theta+i B a_\L L_\L^{1+\zeta}\right]\label{weight} \\
&=&\exp\left[-f L_\L^\theta+i a_\L \left(\frac{L_\L}{\ell_B}\right)^{1+\zeta}\right],\nn
\eea
which differs by the normalizing factor $1+W_\L(0)$ from Eq.~(10) in the main text.
We have defined $\ell_B\equiv  B^{-1/(1+\zeta)}$ and have dropped the explicit dependence of $S_\L$ on $B$. Note that the normalization factor in the denominator in Eq.~(\ref{construction}) ensures that $S_\L(B=0)=1$. Therefore $f L_\L^\theta$ is precisely the free energy difference between the leading and subleading branches of paths, which this model treats as independent from loop energies at smaller scales. 
In weak fields the magnetic field response will be insensitive to the precise value of the small scale cutoff, $L_{\rm min}=N2^{-K}$, as long as it is much smaller than the relevant magnetic length, $L_{\rm min}\ll \ell_B$. Indeed, up to small corrections, $S_\L(B)\approx 1$ for all loops with $L_\L\ll \ell_B$.
 
\subsubsection{2. Non-normalized recursion}
 The above model assumes that free energy differences between a dominant and subdominant branch are independent of the energies (and thus interferences) on smaller scales along those branches. A more realistic model should take into account that if positive interferences occurred along a branch, the resulting "free energy" of the branch is statistically smaller than if the interferences were negligible.
 Such effects can be built into a hierarchical construction by modifying the recursion to   
\bea
S_{\L}&=&1, \quad {\rm if}\quad L_\L= 2^{-K}N,\nn\\
S_{\L}&=& S_{\L'_1}S_{\L'_2}+W_\L(B)  S_{\L''_1}S_{\L''_2},\label{construction_nonnorm}
\eea
 with the same weight factor $W_\L(B)$ (\ref{weight}), but dropping the normalization. In this case $S_{\L}(0)$ is not normalized to $1$ at all length scales. Instead, the explicit contribution to free energy difference between two branches, $f L_\L^\theta$, is now supplemented by an extra contribution coming from the sum over paths at smaller scales. 
 This non-normalized recursion follows a similar hierarchical construction by Derrida and Griffith~\cite{Derrida1989}. Those authors assigned to each loop random energies or signs, that however did not scale with the level of the hierarchy. This generated randomly fluctuating free energies, with a free-energy exponent which is only slightly smaller than the value $\theta=1/3$. In this version of the recursion, we retain the spirit of the Derrida-Griffiths approach with the difference that we introduce the DP-scaling by hand through the free-energy $fL_{\L}^\theta$, and do not consider random signs in the recursion relation.

\subsection{B. Magnetoconductance}
We now study the magnetoconductance of the above models,
\bea
\ln \Delta\sigma_N(B) \equiv  \overline{\ln \left| S_{\L =\{0N\}}(B)\right|}- \overline{\ln \left| S_{\L=\{0N\}}(0)\right|},
\eea
where $\overline{\left[ \cdots \right]}$ denotes the average over the set of reduced free energy and area variables, $\left\lbrace h \equiv (f,a) \right\rbrace $. We assume the two variables associated with each loop to be independent and identically distributed,
\bea
P(\left\lbrace h \right\rbrace) = \prod_\L \rho(f_\L, a_\L)df_\L da_\L.
\eea 
The product runs over all loops $\L$, the support of $\rho$ being $\{f,a\} \in[0,\infty)\times (-\infty,\infty)$. 
However, as we will see, only the values of $\rho(f=0,a)\equiv \rho_a(a)$ will enter the analytical results. 
For quantitative calculations, we will assume a simple Gaussian form, 
\bea
\label{ProbDist}
\rho(f=0,a)\equiv \rho_a(a) = \rho_0 \frac{ \exp[-a^2/2a_0^2] }{\sqrt{2\pi}\,a_0}. 
\eea
The (non-normalized) density $\rho_a(a)$ is a free input parameter of the hierarchical models.
More realistic densities could be determined by studying the distributions of loop areas in the full lattice model.
      
 
Let us now analyze the magnetoconductance,
 \bea
 F(\{h \}) &\equiv& \left[ \ln \left| S_{0N}(B)\right| - \ln \left| S_{0N}(0) \right| \right] ( \{ h\})  
 \label{FDefinition}
 \eea
as a functional of the disorder realization $(\{ h\})$. 
$F$ can be viewed as the free energy difference between a directed polymer with $B$-induced complex weights and one in zero field, where all weights are positive. 

In typical disorder realizations most loops do not play a significant role in modifying the interference of alternative tunneling paths. A loop $\L$ is involved significantly only if $f_\L \lesssim L_\L^{-\theta} $, in which case we refer to it as `active'. 
Large active loops are dilute, while small ones are more abundant, but contribute very little to magnetoconductance.  One can thus expand $F$ in the spirit of a droplet or virial expansion into a sum of terms ${\mathcal V}_k$, which involve an increasing number $k$ of spatially overlapping loops,
 \begin{eqnarray}
 F\left(\left\lbrace h\right\rbrace\right) &=& \mathcal{V}_1 + \mathcal{V}_2 + \mathcal{V}_3 + \dots \nonumber\\
 &=& \sum_{k\geq 1} \sum_{\{\L_1 \neq ...\neq \L_k\}} F^c\left( h_{\L_1},...,h_{\L_k}\right)
 \,. \label{VirialExpansion}
 \end{eqnarray}
The sums are over all (non-ordered) sets of distinct loops.  
The decomposition in Eq.~(\ref{VirialExpansion}) is exact,  given that the connected functions $F^c$ are defined recursively as 
\begin{widetext}
 \begin{align}
 F^c(h_{\L_1}) &= F\left( \left\lbrace h_{\L} \left| f_{\L\neq \L_1} \to \infty  \right.\right\rbrace\right), \label{OneLoopTerm}\\
 F^c(h_{\L_1},h_{\L_2}) &= F\left( \left\lbrace h_{\L} \left| f_{\L \neq \L_1,\L_2} \to \infty  \right.\right\rbrace\right) - F^c(h_{\L_1}) -  F^c(h_{\L_2}), \label{TwoLoopTerms}\\
 \vdots \nonumber\\
  F^c(h_{\L_1},...,h_{\L_k}) &= F\left( \left\lbrace h_{\L} \left| f_{\L}\to \infty\,  \forall \L \not\in \{\L_1,...,\L_k \} \right.\right\rbrace\right) -  
  \sum_{m=1}^{k-1} \sum_{\{\L'_{1} \neq ...\neq \L'_m\}\subset \{\L_{1},...,\L_k\}}
  F^c(h_{\L'_1},... , h_{\L'_m}). \label{DefinitionOfFc}
 \end{align}
 \end{widetext}
 The subtraction of the disconnected terms in Eqs.~(\ref{TwoLoopTerms},\ref{DefinitionOfFc}) ensures that $F^c$ tends to $0$ as one of its free energy arguments becomes large, $f_i\to \infty$, which turns the corresponding loop inactive. It is also easy to verify that $F^c$ vanishes, unless the loops associated with its arguments belong to a single spatially entangled cluster. This follows immediately form the fact that disconnected sets of loops contribute additively to $\ln \left| S_{0N}(B)\right|$. This clustering property ensures an extensive result in the large distance limit, $N\gg \ell_B$  (for every order of the expansion $\overline{{\cal V}_k}\sim N$), i.e., we must have 
\bea
\ln \left| S_{0N}(B)\right| - \ln \left| S_{0N}(0) \right| = -\Delta(\xi^{-1}) N +o(N),
 \eea 
where the coefficient $\Delta(\xi^{-1})$ is expected to be self-averaging. As the notation suggests, this coefficient represents a correction to the inverse localization length $\xi^{-1}$.
 
The disorder average is carried out term by term. Thereby, the disorder variables, especially $f_{\L}$,  take the role of relative positions of particles played in the cluster expansion of gases. The role of a low gas density is played by the small likelihood of large loops to be active. The term ${\cal V}_k$ in the expansion (\ref{VirialExpansion}) captures the interference contribution from \emph{exactly} $k$ active loops, similar to droplet expansions at low $T$ in related disordered systems.~\cite{Doussal2006,Monthus2004,Doussal2010}  
This is akin to the virial expansion, which corrects the ideal gas behavior by summing $n$-particle contributions at order $n^k$ in an expansion in the density $n$. 

The various contributions to ${\cal V}_k$ can easily be represented graphically by enumerating all spatially connected sets of $k$ loops, and summing over their sizes, see Figs. \ref{VirialTerms} and \ref{VirialTermsOrder3} .
 
  
  
\subsection{C. Evaluation of leading terms}

\subsubsection{1. $1^{st}$ order term}

 The first term in Eq.~(\ref{VirialExpansion}) can be rewritten as   

 \bea
 \label{FirstVirialTerm}
   \mathcal{V}_1 &=& \sum_{\L}  \ln \left| \frac{1+W_\L(B)}{1+W_\L(0)}\right| \\
  &=&  \frac{1}{2} \sum_{\L}  \ln \left[ 1 - \frac{4 e^{-f_{\L}L_\L^\theta} \sin^2\left(\frac{a_\L}{2} B L_\L^{1+\zeta} \right)}{\left(1 + e^{-f_{\L}L_\L^\theta}\right)^2} \right], \nn
 \eea
for both the normalized and the non-normalized recursive definitions of the model.

Reorganizing this as a sum over looplengths $\ell_k= N 2^{-k}$, and performing the disorder average, we find
 \bea
 \label{FirstVirialTerm2}
   \overline{\mathcal{V}_1} &=&  \frac{1}{2} \sum_{k=0}^K  \frac{N}{\ell_k}  \overline{\ln \left[ 1 - \frac{\sin^2\left(\frac{a}{2} B \ell_k^{1+\zeta} \right)}{\cosh^2\left(\frac{f}{2} \ell_k^\theta \right)} \right]}^{a,f}\,.
 \eea


\subsubsection{2. $2^{nd}$ order term}
  
\begin{figure}
 \centering
 \includegraphics[angle = 0,width = 0.45 \textwidth]{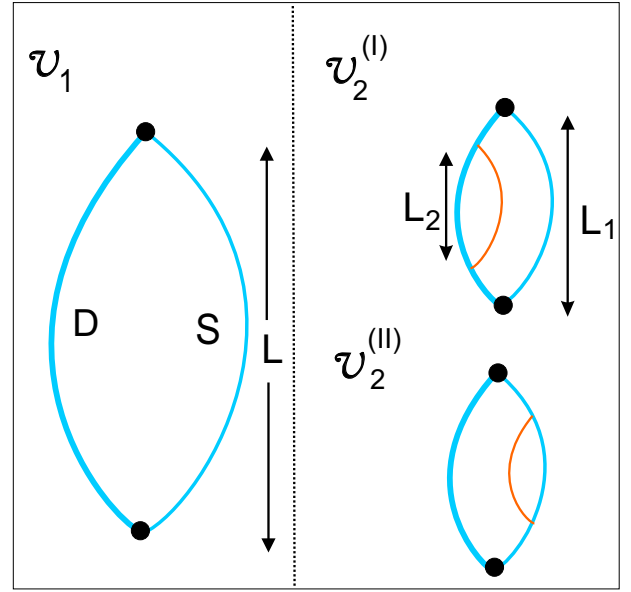}
  \caption{Graphic representation of the first two virial terms in Eq.~\ref{VirialExpansion}. {\em Left:} ${\cal V}_1$ is a sum over all loops $\L$, composed of a dominant (thick line) and subdominant (thin line) branch.  {\em Right}: The two contributions to ${\cal V}_2$ arise form spatially overlapping loops of length $L_1>L_2$. The two cases distinguish whether the smaller loop belongs to the dominant (I) or subdominant (II) branch.}
 \vspace*{-0.2 in}
 \label{VirialTerms}
\end{figure}

The second term in the droplet expansion, ${\cal V}_2$, picks up contributions from disorder realizations where two active loops spatially overlap.  This can occur in two distinct ways, c.f., Fig.~\ref{VirialTerms}: either (I) the smaller loop is part of the dominant;  or (II) part of the subdominant branch of the larger loop.
Let us refer to the bigger and smaller loop as $\L_1$ and $\L_2$, respectively, with lengths $L_{1,2}$. 

The following expressions apply to the normalized model. The discussion of differences for the non-normalized version will be discussed further below when we evaluate the terms.
Denoting $W_{i}(B)\equiv  W_{\L_i}(B)$,
${\cal V}_2$  can be written as 
\begin{eqnarray}
\mathcal{V}_2 = \sum\limits_{\substack{\L_1,\L_2 \\ L_{1}>L_{2}}} \left[\mathcal{V}_2^{(I)}(\L_1,\L_2) + \mathcal{V}_2^{(II)}(\L_1,\L_2)\right]
\label{SecondVirialTerm}
\end{eqnarray}
where 
\bea
\label{SecondVirialTermDiagram1}
 \mathcal{V}_2^{(I)}(\L_1,\L_2) &=& \ln \left| \frac{\frac{1+W_2(B)}{1+W_2(0)}+W_1(B)}{1+W_1(0)}  \right|\\
&&  -\ln \left|\frac{1+W_1(B)}{1+W_1(0)}\right|-\ln \left|\frac{1+W_2(B)}{1+W_2(0)}\right|, \nn
\\
 \mathcal{V}_2^{(II)}(\L_1,\L_2) &=&  \ln \left|\frac{1+\frac{1+W_2(B)}{1+W_2(0)}W_1(B)}{1+W_1(0)}\right|\nn\\
&& -\ln \left|\frac{1+W_1(B)}{1+W_1(0)}\right|.
   \label{SecondVirialTermDiagram2}
\eea
Taking the disorder average and writing (\ref{SecondVirialTerm}) as a sum over loop lengths we have
\begin{eqnarray}
\overline{\mathcal{V}_2} &=& \sum_{k_1=0}^K\sum_{k_2>k_1}^K \frac{N}{2^{k_2}} 
\overline{\ln \left| 1 + \frac{W_1(B) (Z_2 + Z_2^{-1}-2)}{(1+W_1(B))^2} \right|}^{a_{1,2},f_{1,2}}
\label{SecondVirialTerm2}
\end{eqnarray}
where $ Z_2= (1+W_2(B))/(1+ W_2(0))$.


\subsubsection{3. Higher order terms}

For sufficiently large loops, $L_{\L}\gg 1$, the disorder average simplifies. Indeed only very small values of $f_{\L_i}$ are relevant, since the connected functions $F^c$ fall off rapidly when one if its arguments $f_{\L_i}$ is larger than $L_{\L_i}^{-\theta}$. On the other hand, we will see that small scales contribute negligibly to magnetoconductance as long as $B\ll1 $, so we can concentrate on $L_\L\gtrsim \ell_B$. Thus, for each variable $f$  in the disorder average, we  can safely approximate $\int df da \rho(f,a) ... \approx  \int df da \rho_a(a)... $, cf. Eq.~(\ref{ProbDist}). 

Introducing $F_i\equiv  f_{\L_i} L_{\L_i}^\theta$, the disorder-average of the $n$-th virial term becomes

 \bea
 \overline{\mathcal{V}_n} &=& 
 \sum\limits_{\{\mathcal{L}_1\neq ...\neq \mathcal{L}_n\}} \left( \prod\limits_{i=1}^n \frac{1}{L_{\L_i}^\theta}\right) I_n,\label{DisorderAverage} \\ 
 I_n &=&   \prod\limits_{i=1}^n \left( \int_0^\infty dF_i \int_{-\infty}^\infty  \rho_a (a_i) d a_i  \right) F^c(H_1,H_2,\cdots,H_n), \nn
 \eea
using the notation $H_i \equiv (F_i,a_i)$. 

\section{II. Scalings in the droplet expansion}

\subsection{A. Weak fields: $BN^{1+\zeta}\ll 1$}
 
 For weak fields one can expand $I_n$ in the enclosed fluxes, the result being dominated by the largest scale $N$. 
 Expanding Eq.~(\ref{FirstVirialTerm2}) in $B$, and integrating over the rescaled $F$ variables, we find the leading contribution to the magnetoconductance $ \ln \Delta \sigma_N(B)$
\bea
\overline{\mathcal{V}_1} \approx - \frac{1}{4}\int \rho_a(a) a^2 da\, B^2 N \sum_{k=0}^K \ell_k^2 \nn\\
\approx -\frac{1}{3} \int \rho_a(a) a^2 da\, B^2 N^3, \label{weak fields}
\eea
which is negative, as expected for bosonic magnetoconductance.
Likewise, one can check from (\ref{SecondVirialTerm2}), that ${\cal V}_2 \sim O(B^2 N^{2(1+\zeta)-2\theta})$. More generally one finds that higher order terms are suppressed by the prefactors $\prod_{i=1}^n L_i^{-\theta}$ in Eq.~(\ref{DisorderAverage}) with $L_i\sim N$, which leads to the subdominant scaling ${\cal V}_k\sim O(B^2 N^{2(1+\zeta)-k\theta})$. 
 
Note that the leading scaling (\ref{weak fields}) is {\em independent} of the wandering exponent, by virtue of the relation $\theta =2\zeta-1$.  
One therefore obtains the same scaling as in a non-disordered case, for which the exponents $\zeta=1/2, \theta=0$ hold. However, we stress that in the disordered case the result (\ref{weak fields}) arises as a result of disorder averaging, which masks some of the physics.
The {\em distribution} of ${\cal V}_1$ is wide, and the average (\ref{weak fields}) is dominated by a few rare disorder configurations. The latter occur with probability $\sim N^{-\theta}$, but contribute a large ${\cal V}_1\sim B^2 N^{2(1+\zeta)}$, while in most other realizations the wavefunctions are much less affected by quantum interference. \footnote{Since the response in the perturbative regime is strongly inhomogeneous, it is not clear whether the logarithmically disorder-{\em averaged} $\Delta\sigma_N$ with $N=R_{\rm hop}$ is the only relevant quantity determining transport. In particular one should be cautious when using these results as inputs for transport problems on larger scales, such as variable range hopping. We are not aware of any theoretical approach which take into account the statistical distribution of the $B$-effects on wavefunction properties, rather than assuming a homogeneous average effect on all wavefunctions.}

  \begin{figure}
 \centering
 \includegraphics[angle = 0,width = 0.45 \textwidth]{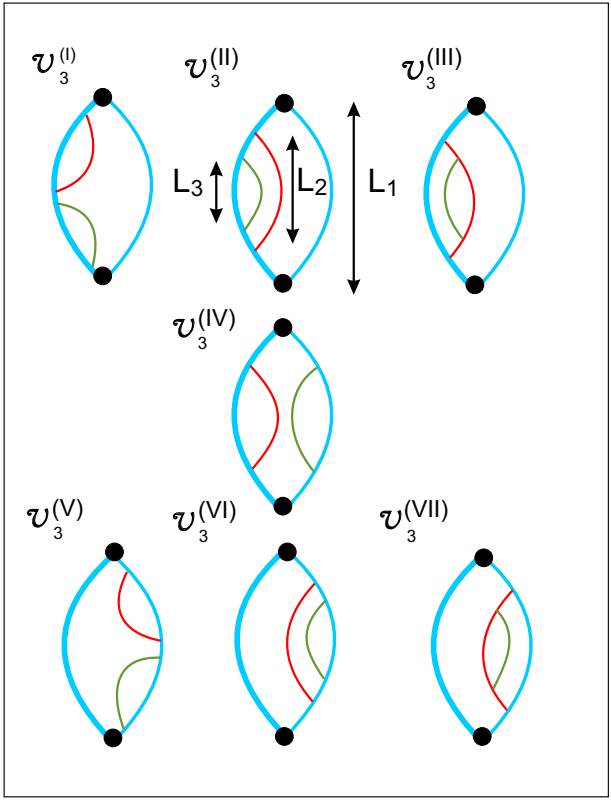}
  \caption{Graphic representation of the third order terms ${\cal V}_3$ in Eq.~(\ref{VirialExpansion}) with 
 $ L_1 $ (blue) $\geq L_2$ (red) $\geq L_3$(green).}
 \vspace*{-0.2 in}
 \label{VirialTermsOrder3}
\end{figure}

 \subsection{B. Strong fields: $BN^{1+\zeta}\gg 1$}
 The perturbative expansion holds only for weak fields for which the distance between end points is smaller than the `magnetic length', $N \ll \ell_B$. For stronger fields, the dominant contribution comes from loops at the scale $\ell_B$. To see this, let us approximate the sum over discrete loop sizes in (\ref{FirstVirialTerm2}) as an integral, $\sum_k\approx 1/\ln(2)\int d\ell/\ell$,
 \bea
 \label{FirstVirialTerm3}
   \overline{\mathcal{V}_1} 
   &\approx &  \frac{N}{2\ln(2)} \int_1^N  \frac{d\ell}{\ell^2} \, \overline{\ln \left[1 - \frac{\sin^2\left(\frac{a}{2} B \ell^{1+\zeta} \right)}{\cosh^2\left(\frac{f}{2} \ell^\theta \right)} \right]}^{a,f}\,.
 \eea
Rescaling the free energies and changing variables to  $u\equiv \ell/\ell_B$, we obtain 
 \bea
 \label{FirstVirialTerm3Repeat}
   \overline{\mathcal{V}_1} 
   &= & -c_1\frac{N}{\ell_B^{1+\theta}} = -c_1 NB^{\frac{2\zeta}{1+\zeta}} = -c_1 N B^{4/5},
   \eea
   with the numerical coefficient
   \bea
   c_1 &\approx & -\frac{1}{2\ln(2)} \int_0^\infty \frac{du}{u^{2+\theta}} \int_0^\infty dF \times\nn\\
   && \quad \int_{-\infty}^\infty da \rho_a(a) \ln\left[1-\frac{\sin^2(a u^{1+\zeta}/2}{\cosh^2(F/2)}\right].
 \eea
 For the particular choice  (\ref{ProbDist}) for $\rho_a(a)$ (with $\rho_0=a_0=1$) we find 
 \bea
 c_1\approx 0.86.
 \eea 
Note that the dominant contribution comes indeed from $u=O(1)$, i.e., from loops of size $\ell_B$. We have extended the limits of the $u$-integral to $0$ and $\infty$, as it converges rapidly on both sides.
This result implies a leading correction to the localization length as
\bea
\label{dxi}
\Delta(\xi^{-1}) = c_1 B^{\frac{2\zeta}{1+\zeta}}.
\eea

\subsection{C. Subleading corrections}

In the non-perturbative regime, subleading corrections are interesting to analyze in more detail, as they correct the leading behavior (\ref{dxi}). As we shall see below, there is a direct correlation between the order of a term in the virial expansion and its scaling with $B$, which justifies using the virial expansion in the first place. Roughly speaking, each loop contributes a scaling factor of $B^{\frac{\theta}{1+\zeta}}$ on disorder averaging, causing the $n$-th virial term with $n$ spatially overlapping loops to contain as many such factors.

A more precise formulation follows. We begin with the second order term $\mathcal{V}_2$.
As before, we can rewrite $\overline{\mathcal{V}_2}$ as a sum over pairs of loop lengths, $L_j = 2^{-j} N$ and $L_k = 2^{-k} N $, and take the continuum limit of the discrete sums over $j$ and $k$
\bea
  \overline{\mathcal{V}_2} &=& \sum\limits_{\substack{j,k \\k > j}} \frac{N}{L_{k}} \left[ \overline{\mathcal{V}_2^{(I)}}(L_{j},L_k) + \overline{\mathcal{V}_2^{(II)}} (L_{j},L_{k})\right] \label{SecondVirialTermAvg} \\
  &\approx& \frac{N}{(\ln 2)^2}\int_1^N \frac{d\ell_1}{\ell_1}  \int_1^{\ell_1} \frac{d\ell_2}{\ell_2^2}\left[ \overline{\mathcal{V}_2^{(I)}}(\ell_1,\ell_2) + \overline{\mathcal{V}_2^{(II)}} (\ell_1,\ell_2)\right]. \nonumber
\eea
 Here, $\mathcal{V}_2^{(I)}\left(\ell_1,\ell_2\right) $ and $\mathcal{V}_2^{(II)}\left(\ell_1,\ell_2\right) $ are given by Eqs.~(\ref{SecondVirialTermDiagram1},\ref{SecondVirialTermDiagram2}) (loop $\L_i$ being referred to by its length $\ell_i$). The overbar denotes the average over $f_{1,2}$ and $a_{1,2}$ with the appropriate probability distributions
 \begin{multline}
  \overline{\left[ \cdots\right]} \equiv \int_{0}^\infty df_1 \rho_f(f_1) \int_0^\infty df_2 \rho_f(f_2) \\ \int_{-\infty}^\infty da_1 \rho_a(a_1) \int_{-\infty}^\infty da_2 \rho_a(a_2) \left[ \cdots \right], \nonumber
  \end{multline} 

%
Substituting $u_1 = \frac{\ell_1}{\ell_B}$ and $u_2 = \frac{\ell_2}{\ell_B}$, we obtain the subleading correction 
  \bea
  \overline{\mathcal{V}_2} &=& -c_2 \frac{N} {\ell_B ^{(1 + 2\theta)}}  = -c_2 NB^{\frac{4\zeta - 1}{1+\zeta}}  ,
  \eea
  with the numerical coefficient 
  \begin{widetext}
  \begin{multline}
 c_2 \approx -\frac{1}{(\ln 2)^2}  \int_0^\infty \frac{du_1}{u_1^{1+\theta}} \int_0^{u_1} \frac{du_2}{u_2^{2+\theta}} \int_0^\infty dF_1 \int_0^\infty dF_2 \int_{-\infty}^\infty da_1 \rho_a(a_1) \int_{-\infty}^\infty da_2 \rho_a(a_2) \\
  \times \ln \left| 1 + \frac{ e^{-F_1 + i a_1 u_1^{1+\zeta}}}{\left(1 +e^{-F_1 + i a_1 u_1^{1+\zeta}}\right)^2} \left(\frac{1+e^{-F_2 + i a_2 u_2^{1+\zeta}}}{1 + e^{-F_2}}  +\frac{1 + e^{-F_2}}{1+e^{-F_2 + i a_2 u_2^{1+\zeta}}}-2 \right)\right|.
   \label{c2}
  \end{multline}
  \end{widetext}

Note that the integrals converge both for $u_{1,2}\to 0$ and $u_{1,2}\to \infty$.

We have computed the $(F_1,F_2,a_1,a_2)$-integral in Eq.~(\ref{c2}) using Monte-Carlo sampling, followed by numerically carrying out the $(u_1,u_2)$-integration. With the density $\rho_a$ given in (\ref{ProbDist}), $c_2$ turns out to be negative. 

To the first subleading order we find the correction to the inverse localization length as 
\bea
\Delta(\xi^{-1}) = -c_1 B^{4/5}\left(1+\frac{c_2}{c_1}B^{1/5}+O(B^{2/5})\right). 
\eea
Note that the subleading corrections vary slowly with $B$ and thus are expected to affect fits of the magnetoconductance to a simple power law $B^\gamma$. Indeed defining an `effective exponent' 
as 
\bea
\!\!\!\gamma = \frac{d \ln|\Delta(\xi^{-1})|}{d\ln B} =   \frac{4}{5}+\frac{1}{5}\frac{B^{1/5}}{c_1/c_2+ B^{1/5}}+O(B^{2/5}),\,
\eea
one expects to see apparent exponents that deviate from the asymptotically exact value 4/5 for any small but finite $B\ll 1$.
The sign of the correction depends on the relative sign of $c_1$ and $c_2$.

The numerical data obtained for the full lattice model is consistent with a positive correction to the exponent, cf. inset of Fig. 3 in main text..
However, the normalized hierarchical model predicts the opposite sign. We believe that this qualitative difference is due to the fact that the normalized recursion neglects correlations of free energy differences at different scales, as explained above.

A more realistic model, which builds in such correlations was given in Eq.~(\ref{construction_nonnorm}), where the normalizing factors are  dropped in the recursive definition of path weights. The expression for ${\cal V}_2$ is easy to derive in this case as well,  

  \begin{widetext}
  \bea
 c_2 &\approx& - \frac{1}{(\ln 2)^2}  \int_\Lambda^\infty \frac{du_1}{u_1^{1+\theta}} \int_\Lambda^{u_1} \frac{du_2}{u_2^{2+\theta}} \int_1^\infty dF_1 \int_0^\infty dF_2 \int_{-\infty}^\infty da_1 \rho_a(a_1) \int_{-\infty}^\infty da_2 \rho_a(a_2) \left[ \mathcal{V}_2^{(I)} + \mathcal{V}_2^{(II)} \right],   \label{c2modified}
 \\
 \mathcal{V}_2^{(I)} &=& \ln \left| \frac{1 +  e^{-F_1 + ia_1 u_1^{1+\zeta}} + e^{-F_2 + ia_2 u_2^{1+\zeta}}}{1+e^{-F_1} + e^{-F_2}}\right| -\ln \left| \frac{1 + e^{-F_1 + ia_1 u_1^{1+\zeta}}}{1+e^{-F_1}}\right| - \ln \left| \frac{1 + e^{-F_2 + ia_2 u_2^{1+\zeta}}}{1+e^{-F_2}}\right|, \nonumber \\
 \mathcal{V}_2^{(II)}  &=& \ln \left| \frac{1 +  e^{-F_1 + ia_1 u_1^{1+\zeta}}(1 + e^{-F_2 + ia_2 u_2^{1+\zeta}})}{1+e^{-F_1}(1 + e^{-F_2})}\right| -  \ln \left| \frac{1 + e^{-F_1 + ia_1 u_1^{1+\zeta}}}{1+e^{-F_1}}\right|, \nonumber
  \eea
  \end{widetext}
where the lower cutoff $\Lambda$ is a number $\lesssim 1$, ensuring that the recursion ends at $L_k = \Lambda \ell_B$. We used $\Lambda = 0.125$ for the numerical computation of $c_2$ below. This cutoff is required since the $u_2$-integral in Eq.~(\ref{c2modified}) does not converge at infinitesimally small length-scales. 
This reflects the fact that the $S_{\L}$ (and thus the loop free energies) have a non-trivial distribution already in $B=0$, due to interferences at small scales $L\ll \ell_B$. This distribution cannot be captured easily by the virial expansion. Instead we have to introduce a small scale cut-off at some fixed length scale $L_k \lesssim \ell_B$. We can safely assume that the small scale interference is incorporated into the free energy differences at that smallest scale. Thereby we rely on the fact that  smaller loops enclose negligible flux and thus do not contribute significantly to magnetoconductance, nor affect much the free energy distribution at small scales. 
Finally, this prescription leads to a similar virial expansion in powers of $B^{1/5}$, however with different coefficients $c_{k>1}$.  

   
The integral~(\ref{c2modified}) yields $ c_2 \approx 4.9 \times 10^{-2}$. This has the {\em same} sign as $c_1$and thus leads to an ``effective exponent" which is bigger than 4/5, and thus comes closer to the phenomenology observed in the full lattice model, as one may expect. As mentioned before, the various definitions of the hierarchical construction only affect the coefficients of the {\em subleading} terms in the virial expansion.

\subsection{D. Effect of small denominators and resonances}
The quantitative effect of the subleading terms is of course non-universal, as are the coefficients $c_{1,2}$. 
A variation of such effects is actually also found in the full lattice sum of forward-directed paths. It may seem dangerous to evaluate path sums of products of denominators which can become arbitrarily small. While the logarithmic average of such sums is mathematically well-defined, it is known that backscattering and self-energy effects, or a Coulomb gap in the density of states, reduce the influence of such resonances. For this reason the toy models considered in the earlier literature~\cite{Shklovskii1991,Prior2009} have restricted themselves to finite denominators.

Numerically evaluating the sum over all paths as given in the maintext, without restricting the occurrence of resonant denominators, we found effective exponent of the order of $\gamma\approx 0.88$. 
However, the deviation from $4/5$ turned out to be much smaller for a toy model where we restricted onsite energies to the interval $[1/2,1]$. 
It is thus suggestive to attribute the stronger deviations with resonances included to an enhanced value of $c_2$. 
 
\subsection{E. Higher terms in the droplet expansion} 
It is not difficult to write down the disorder-average of the higher order terms in Eq.~\ref{VirialExpansion} as appropriate integrals. One can check that the generic term $\overline{\mathcal{V}_k}$ varies as
 \bea
 \overline{\mathcal{V}_k} = c_k N B^{\frac{1+k\theta}{1+\zeta}}
 \eea 

To illustrate the procedure, we give the diagrams contributing to ${\cal V}_3$ in Fig. \ref{VirialTermsOrder3}. The corresponding expressions for the connected terms are given below. Subscripts 1, 2 and 3 denote three loops with lengths $L_1 \geq L_2 \geq L_3$. For brevity, we only consider for the normalized model and give the {\em connected} terms in $\mathcal{V}_3^{(k)}$ as $ V_3^{(k)}(B) - V_3^{(k)}(B=0)$, where
\bea
V_3^{(I)}(B) &=& \ln \left| 1 + W_1 + W_2 + W_3 + W_2 W_3 \right| ,\nn  \\
V_3^{(II)}(B) &=& \ln \left| 1 + W_1 + W_2 + W_3\right| ,\nn\\
V_3^{(III)}(B) &=& \ln \left| 1 + W_1 + W_2 + W_2 W_3\right|,\nn \\
V_3^{(IV)}(B) &=& \ln \left| 1 + W_1 + W_2 + W_1 W_3  \right| ,\nn\\
V_3^{(V)}(B) &=& \ln \left| 1 + W_1 + W_1 W_2 + W_1W_3 + W_1W_2W_3\right| ,\nn \\
V_3^{(VI)}(B) &=& \ln \left| 1 + W_1 + W_1 W_2 + W_1 W_3\right| ,\nn \\
V_3^{(VII)}(B) &=& \ln \left| 1 + W_1 + W_1 W_2 + W_1 W_2 W_3\right|, \nn 
\eea 
where $ W_1 \equiv W_{\L_1}(B)$.
Continuing along these lines,  $\ln \Delta\sigma_N(B) $ can be calculated to any desired order at a given field $B$.

\subsection{F. Remarks on fermions}

It might be interesting to  generalize the hierarchical model to the case of fermions. Since the locator expansion yields path amplitudes with positive and negative signs, it would seem natural to include random signs $s_\L$ in a hierarchical droplet model. However, several subtleties may need further modifications to capture the details of fermionic magnetoconductance. For example, a weak field can have a significant effect on small loops whose branches have nearly opposite amplitudes. This may reflect in a non-trivial dependence of  free energy costs $f_\L$ on $B$, which may enhance subleading corrections and potentially even change their exponent. It is possible that the observed effective fermionic exponents $\gamma<4/5$ in the non-perturbative regime are due to such effects. More detailed investigations are necessary to clarify these issues.

\end{document}